# Effect of Pb doping on the crystallization process and thermoelectric properties of Ge$_2$Sb$_2$Te$_5$ phase change material

M. Zhezhu[1*], A. Vasil'ev[1], M. Yaprintsev[2], A. Musayelyan[3], E. Pilyuk[2], O. Ivanov[2]

[1]A.B. Nalbandyan Institute of Chemical Physics NAS RA, Yerevan 0014, Armenia
[2]Belgorod State University, Belgorod 308000, Russia
[3]Institute of Radiophysics and Electronics NAS RA, Ashtarak 0204, Armenia

[*]Author to whom correspondence should be addressed to marina.zhezhu@ichph.sci.am

**Abstract**

**Phase change materials based on Ge–Sb–Te alloys are widely explored for their potential in both memory devices and thermoelectric applications. In this study, films of Ge$_2$Sb$_2$Te$_5$ (GST) doped with varying concentrations of Pb were prepared and systematically investigated to trace the effect of Pb doping on crystallization-induced phase transformations and thermoelectric properties. *Via* X-ray diffraction and Raman spectroscopy, the impact of Pb doping on the crystallization behavior was revealed and examined. According to the specific electrical resistivity measurements, the Pb doping resulted in decreasing both the amorphous-to-cubic and cubic-to-hexagonal transition temperatures, thereby facilitating the formation of the hexagonal phase at a lower thermal regime. Furthermore, Seebeck coefficient and electrical resistivity data of hexagonal Pb-doped GST were used to calculate the power factor, *PF*. A *PF* maximum equal to 1.3 mW/(K$^2$·m) was found for 2.5 at.% Pb-GST at 633 K, with the highest carrier mobility also observed for this composition ($\mu_H$ = 92.0 cm$^2$/(V·s)). Controlled Pb doping effectively modulates both structural transitions and thermoelectric performance, highlighting the potential of Pb–GST for applications that combine phase-change memory and thermoelectric functionality, such as opto-thermoelectric devices and non-volatile thermoelectric sensors.**

## 1. Introduction

The constant demand for faster data storage, energy-efficient electronic devices, and advanced thermoelectric materials continues to drive the search for functional materials with tunable structural and transport properties [1, 2]. Among them, phase-change materials (PCMs) occupy a central position due to their ability to switch reversibly between distinct structural states, making them indispensable for non-volatile memory technologies while also positioning them as promising candidates for emerging thermoelectric applications [3-6]. Moreover, combined phase change and thermoelectric properties in a single material open exciting opportunities for specific memory devices [7], next-generation energy technologies [8, 9], on-chip thermoelectric generators [10], and even ocean power applications [11].

One of the most studied PCMs is $Ge_2Sb_2Te_5$ (GST), which not only serves as a benchmark for phase-change memory but also shows promise for thermoelectric applications [12-15]. GST can exist in three distinct structural forms: the amorphous state, the metastable face-centered cubic (*fcc*) phase, and the stable hexagonal close-packed (*hcp*) phase. Amorphous GST is characterized by a highly disordered atomic network with broken chemical order and a large number of vacancies, resulting in high electrical resistivity and low carrier mobility [5, 15]. Upon heating (about 125°C to 175°C [15-17]), it transforms into the metastable cubic phase, in which atoms arrange into a rocksalt-like structure with significant vacancy disorder, enabling fast crystallization—a key feature for phase-change memory [18]. Further annealing (250 °C ÷300 °C [17, 19, 20]) leads to the stable hexagonal phase, which exhibits higher atomic order, reduced vacancy concentration, and improved electronic transport [5]. In the hexagonal phase, GST films also demonstrate enhanced thermoelectric properties, including higher mobility and Seebeck coefficient [5], making it promising for multifunctional devices combining memory and energy-conversion capabilities.

However, the thermoelectric properties of GST-based films are rarely assessed in conjunction with their phase transitions. To date, no systematic study has examined how doping can simultaneously modulate phase transitions and influence the thermoelectric performance of these materials. Doping strategies provide a promising route to tailor the properties of GST. By substituting host atoms with dopants, it is possible to modify lattice parameters, adjust carrier concentrations, and stabilize structural states [15]. Several doped GST films have been extensively investigated for data storage applications, demonstrating notable improvements such as enhanced thermal stability with As doping [21] and Sm doping [22], increased resistance contrast and one-step crystallization with Se doping [19], raised transition temperatures with In doping [23], decreased crystallization temperature with Sn doping [24], *etc*. In addition, doped GST materials have also been investigated as promising thermoelectrics with studies exploring the impact of doping on band structure, scattering mechanisms, and carrier concentration [5, 14, 25].

Nevertheless, Pb doping or substitution in GST remains largely unexplored. Compared to Ge, Pb acts as an isovalent dopant and could impart unique modifications to the structural and electronic properties of the material. In particular, Pb doping emerges as a promising strategy to adjust the behavior of GST. Previous first-principles studies of Pb-doped GST have shown enhanced phase-change properties [26]. While in thermally evaporated Pb:GeSbTe chalcogenide films, an increase in the amorphous-to-crystalline transformation temperature was observed [27]. Moreover, in the context of thermoelectricity, Pb doping has been shown to significantly influence the thermoelectric performance [28-31].

In this work, we systematically investigated the effect of Pb doping on the crystallization process and thermoelectric properties of GST films. $Ge_2Sb_2Te_5$ films with Pb concentrations of 2.5, 4.8, and 6.8 at.% were deposited by magnetron sputtering. Structural and vibrational evolution across the amorphous, metastable cubic, and stable hexagonal phases was analyzed using X-ray diffraction and Raman spectroscopy. Temperature-dependent resistivity measurements were performed to track crystallization pathways; meanwhile, the thermoelectric performance of the hexagonal phase was assessed through the

Seebeck coefficient, resistivity, and power factor. Carrier mobility measurements were further used to evaluate charge transport efficiency and clarify the role of Pb incorporation. By elucidating how Pb doping affects both structural transitions and charge transport, this study provides strategies for tailoring phase behavior and optimizing performance in GST-based energy and functional materials.

## 2. Materials and Methods

Pb-doped $Ge_2Sb_2Te_5$ films were prepared *via* RF magnetron co-sputtering using separate PbTe and $Ge_2Sb_2Te_5$ alloy targets. Undoped GST was also prepared as a reference sample. For GST target preparation, Ge, Sb, and Te powders were thoroughly mixed in stoichiometric proportions. The mixed powders were consolidated *via* spark plasma sintering (SPS) at 400 °C under a pressure of 40 MPa for 5 min, resulting in a cylindrical target with dimensions of Ø40 mm × 15 mm. The PbTe target was prepared using the same SPS procedure by mixing Pb and Te powders in a stoichiometric ratio, yielding a cylindrical target of Ø10 mm × 15 mm. The films were deposited onto glass substrates without additional heating. The background pressure was $1.9 \times 10^{-2}$ Pa, and Ar gas was introduced at 0.3 Pa during deposition. The RF power on the PbTe target ranged from 3 to 13 W, while the RF power on the GST target was 50 W. A schematic representation of the sputtering process for the films is shown in Fig. S1.

The surface morphology, thickness (averaged over 20 measurement points across the film), and elemental composition (determined by energy-dispersive spectroscopy, EDS) were characterized using scanning electron microscopy (SEM, Nova NanoSEM 450). EDS measurements were carried out at 10 kV, 1812× magnification, 3.84 μs amplifier time, and 128 eV energy resolution. For major elements, the relative compositional precision is ±1%. The chemical composition and valence states of the elements were analyzed using K-Alpha X-ray photoelectron spectroscopy (XPS, Thermo Fisher Scientific). The phase structures of the Pb-doped GST films were analyzed by X-ray diffraction (XRD, Rigaku Ultima IV diffractometer with CuKa-radiation) in the *2θ* range from 10° to 80°. Raman analysis was performed over the range 40-350 cm$^{-1}$ using a LabRam HR Evolution (Horiba) equipped with a 600 gr/mm grating (200 μm hole) and a 5mW laser. For X-ray diffraction and Raman analysis, the as-deposited films with different Pb-doping concentrations were subsequently annealed in an inert argon atmosphere (99.998%) at 225 °C and 325 °C for 30 minutes each, followed by slow cooling to room temperature. This treatment induced the phase transition from amorphous to *fcc* phase at 225 °C and from *fcc* to *hcp* at 325 °C. The Rietveld refinement was applied to calculate crystal lattice parameters. Electrical properties of the Pb-doped GST films were measured using a ZEM-3 system (ULVAC, Inc.). The measurement accuracy of the system is approximately ±5% for the Seebeck coefficient and ±3% for electrical resistivity. The Ecopia HMS-5000 was used to determine the electrical transport properties.

## 3. Results and Discussion

### 3.1 Structural and Phase Evolution of Pb-Doped GST Films

Since the crystallization of amorphous films occurs through localized regions, the phase transition of the entire film is temporally extended. Moreover, the stability ranges of the metastable *fcc* and stable *hcp* phases span broad temperature intervals. Therefore, to reliably evaluate the properties of a given phase, it is necessary to ensure that crystallization is as complete as possible. Previous studies on the temperature-dependent structural evolution of $Ge_2Sb_2Te_5$ reported the formation of the *fcc* phase after annealing at 180 °C and the *hcp* phase at 370 °C [32]. In addition, investigations of the effect of doping on crystallization temperature reported the onset of *fcc* crystallization in the range of 160–275 °C [15] and the formation of the *hcp* phase between 300 °C and 350 °C [33, 34]. Consistent with these reports, annealing temperatures of 225 °C (for amorphous-to-*fcc* transition) and 325 °C (*fcc*-to-*hcp* transition) were selected as representative values to achieve fully developed crystalline phases. The temperature dependence of the electrical resistivity presented in Section 3.2 further supports the appropriateness of these chosen annealing temperatures for the amorphous-to-*fcc* and *fcc*-to-*hcp* phase transitions.

The as-deposited amorphous films exhibited a smooth and homogeneous morphology with no visible surface defects (see Fig. S1). The film thicknesses and chemical compositions are presented in Table 1. The GST films exhibited a Ge:Sb:Te stoichiometry of 2:2:5 and included undoped and Pb-doped compositions, with lead concentrations of 2.5 at.%, 4.8 at.%, and 6.8 at.%, referred to as 2.5 at.% Pb-GST, 4.8 at.% Pb-GST, and 6.8 at.% Pb-GST, respectively. Such Pb concentrations fall within the composition ranges that have previously been identified as particularly relevant for tuning the structural and electronic properties of telluride-based PCMs. First-principles calculations have indicated that a Pb content of approximately 4.4 at.% corresponds to a thermodynamically stable substitutional doping level in GST [26]. In addition, experimental studies on Pb-containing $Ge_2Sb_2Te_5$ compositions have examined Pb atomic contents in the range of 1–5 at.% and have demonstrated that Pb incorporation within this range significantly affects phase transformation behavior [27]. Furthermore, even low Pb doping levels ($0.005 \leq x \leq 0.0125$ in $Sb_{2-x}Pb_xTe_3$) have been reported to noticeably modify carrier transport properties [29]. Therefore, the Pb concentrations (2.5–6.8 at%) used in this study fall within a compositional range that has previously been identified as especially relevant for influencing the structural and electronic characteristics of related PCMs.

Table 1. Summary of the thicknesses and chemical compositions of GST and Pb-doped GST films, as well as their lattice parameters after annealing at 225 °C (cubic *fcc* phase) and 325 °C (*hcp* phase).

| Film | Thickness, nm | Chemical composition, at. % | | | | Lattice parameters of annealed films | | |
|---|---|---|---|---|---|---|---|---|
| | | Pb | Ge | Sb | Te | cubic *fcc* phase | hexagonal *hcp* phase | |
| | | | | | | a, Å | a, Å | c, Å |
| GST | 279 | 0 | 25.4 | 23.3 | 51.4 | 6.007 | 4.253 | 18.355 |
| 2.5 at. % Pb-GST | 202 | 2.5 | 23.8 | 22.2 | 51.6 | 6.0889 | 4.2767 | 18.121 |
| 4.8 at. % Pb-GST | 205 | 4.8 | 21.7 | 20.7 | 52.7 | 6.098 | 4.272 | 18.139 |
| 6.8 at. % Pb-GST | 155 | 6.8 | 21.4 | 20.2 | 51.5 | 6.214 | 4.264 | 18.168 |

Fig. 1 (a-c) presents the XRD patterns of films and shows the structure evolution by annealing. As-deposited films showed a mountain peak at 28.3°, indicating that the prepared films are amorphous [35]. After annealing at 225 °C, the films crystallize into the metastable *fcc* phase with the space group Fm-3m (PDF Card No.: 01-085-3193). In the Pb-doped films (2.5, 4.8, and 6.8 at.% Pb), the diffraction peaks shift toward lower angles, indicating an increase in the interplanar spacing (*d*) and suggesting lattice expansion. This behavior can be attributed to the larger covalent radius of Pb (146 pm) compared to Sb (139 pm) and Ge (120 pm) [36]. The calculated lattice parameters for these compositions are presented in Table 1. The introduction of Pb increases the lattice spacing of the *fcc* phase, in agreement with experimental observations for isovalent doping in GST films [37].

Upon further annealing at 325 °C, the films undergo a phase transition to a *hcp* phase with the space group P-3m1 (PDF Card No.: 01-082-8885). For samples containing 4.8 and 6.8 at. % Pb, additional diffraction peaks appear at $2\theta$ = 39.65° and 49.9°, corresponding to the (220) and (222) planes of the hexagonal PbTe phase (PDF Card No.: 01-075-8255), indicating the formation of a secondary phase at higher Pb concentrations. Due to the small thickness of the films (~200 nm) and partial peak overlap, a reliable quantitative determination of the PbTe volume fraction from conventional XRD data is not feasible. Trial Rietveld refinements, including a PbTe phase, yielded unstable scale factors with large parameter correlations, indicating that the available data do not support a reliable quantitative analysis. Nevertheless, a rough estimate based on peak intensity ratios suggests that the PbTe fraction does not exceed approximately 1–1.5 vol.% for both samples with Pb concentrations of 4.8 and 6.8 at.%, although this value should be treated with caution due to the limitations mentioned above.

Furthermore, the increased peak broadening observed in Pb-doped films (see insets of Fig. 1(b-c)) suggests local lattice distortions caused by the incorporation of Pb atoms, compared to undoped GST, consistent with previous reports on doped GST films [19, 37].

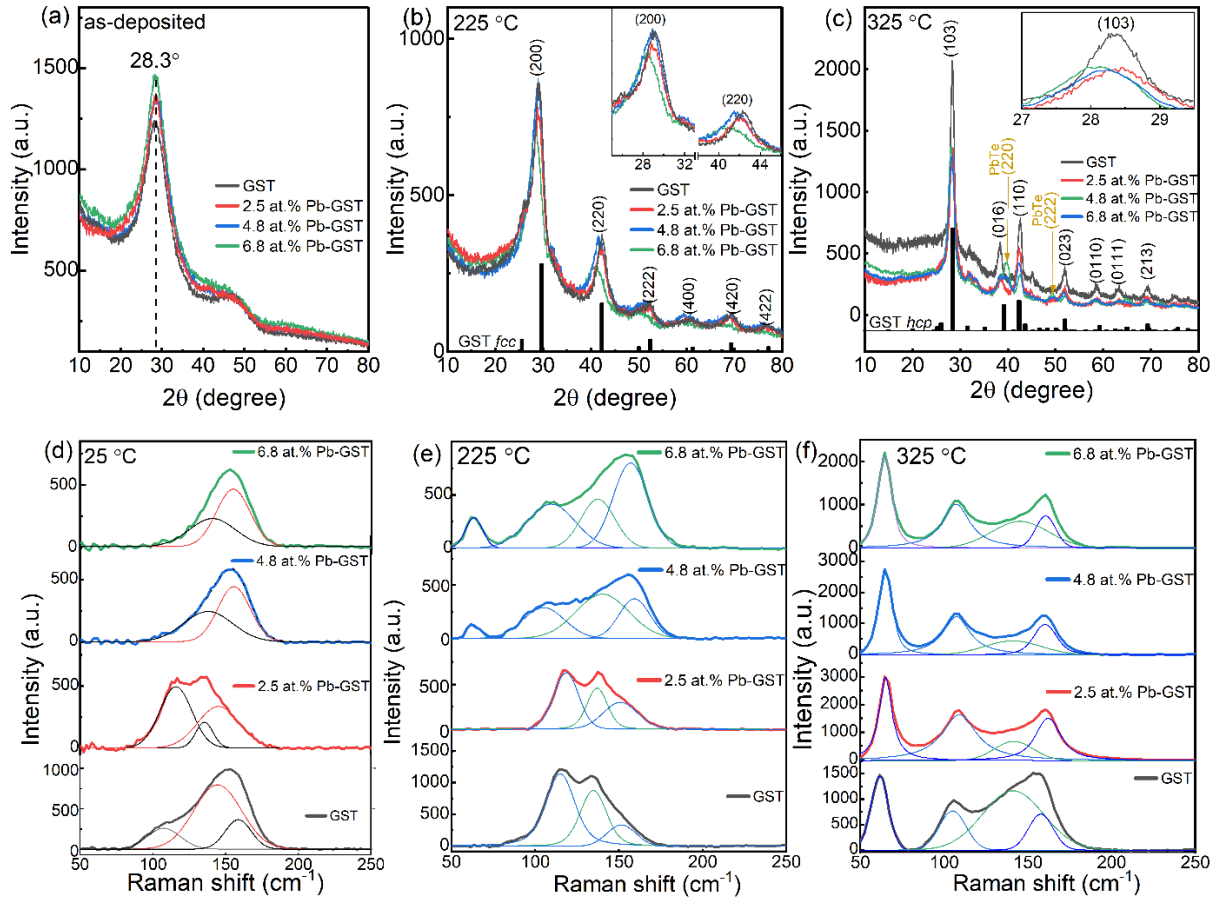

Fig. 1. XRD patterns of Pb-doped GST films: (a) as-deposited, (b) after annealing at 225 °C, and (c) after annealing at 325 °C. Insets in (b) and (c) show zoomed-in peaks. Raman spectra of Pb-doped GST films: (d) as-deposited, with the main peak highlighted in red; (e) after annealing at 225 °C, with the GeTe-related peak highlighted in green and the Sb-Te-related peaks in blue; (f) after annealing at 325 °C, with the GeTe-related peak in green and the Sb-Te-related peaks in blue.

Analysis of the lattice parameters of the hexagonal phases (Table 1) revealed that parameter $a$ slightly increases, while parameter $c$ decreases with increasing Pb content in the films. According to first-principles calculations reported in [26] for 4.4 at. % Pb-doped GST, a decrease in equilibrium volume occurs after doping. Although Pb substitution at the Ge site does not significantly alter the Te–Ge and Te–Sb bond lengths, it leads to a shortening of Te–Te bonds and a reduction in the $c$ parameter. This correlation suggests that Pb incorporation introduces local lattice distortions, promoting structural densification and contributing to the observed anisotropic changes in lattice parameters.

Raman spectroscopy was employed to analyze structural evolution induced by Pb doping. Figure 1 (d–f) shows the spectra of the as-deposited sample at room temperature, at 225 °C, and at 325 °C, in correlation with the XRD measurements. The corresponding Gaussian-fitted Raman peak positions are summarized in Table S1. As shown in Fig. 1(d), the Raman spectra of amorphous GST and 2.5 at.% Pb–GST consists of three sub-peaks. The dominant sub-peak of undoped GST at 144.6 cm$^{-1}$ is associated with amorphous GST and closely resembles the spectrum of amorphous $Sb_2Te_3$ [19, 38, 39]. An increase of Pb content leads to a blueshift of the main peak, indicating changes in chemical bond lengths and angles.

Detailed fitting of the Raman spectra for the films in the *fcc* phase is shown in Fig. 1(e). Both undoped GST and 2.5 at.% Pb-GST exhibit three characteristic subpeaks. The main subpeak, appearing at 115.0 cm$^{-1}$ for GST and at 118.5 cm$^{-1}$ for 2.5 at.% Pb-GST corresponds to the $E^2_g$ mode of $Sb_mTe_3$ ($m$ = 1, 2) [40-42]. A lower subpeak assigned to the $A_1$ mode of edge-sharing $GeTe_{4-n}Ge_n$ ($n$ = 0–3) tetrahedra [42] is observed at 134.8 cm$^{-1}$ for GST and at 137.0 cm$^{-1}$ for 2.5 at. % Pb-GST. The smallest subpeak at 151.5 cm$^{-1}$ and 150.9 cm$^{-1}$ for GST and 2.5 at.% Pb-GST, respectively, is attributed to the $A^2_{1g}$ mode of $Sb_mTe_3$ ($m$ = 1, 2) [40, 41]. In contrast, fitted Raman spectra of 4.8 and 6.8 at.% Pb-GST in the *fcc* phase demonstrate four subpeaks. The first peak, observed at 63.1 cm$^{-1}$ and 63.5 cm$^{-1}$ for 4.8 at.% and 6.8 at.% Pb-GST, respectively, corresponds to the $A^1_{1g}$ vibrational mode of $Sb_mTe_3$ ($m$ = 1, 2) [40]. The second peak, at 105.1 cm$^{-1}$ and 109.6 cm$^{-1}$, is assigned to the $E^2_g$ mode of $Sb_mTe_3$ ($m$ = 1, 2) [40-42]. The third peak, located at 140.2 cm$^{-1}$ and 137.5 cm$^{-1}$, is associated with the $A_1$ mode of edge-sharing $GeTe_{4-n}Ge_n$ ($n$ = 0–3) tetrahedra [42], while the fourth peak, at 159.3 cm$^{-1}$ and 157.0 cm$^{-1}$, corresponds to the $A^2_{1g}$ mode of $Sb_mTe_3$ ($m$ = 1, 2) unit [40, 41]. Hence, in the films with higher Pb concentrations (4.8 and 6.8 at.%), Sb-Te associated vibrational modes become more pronounced, evidenced by the appearance of the $A^1_{1g}$ vibrational mode and the dominance of the $A^2_{1g}$ mode of $Sb_mTe_3$ ($m$ = 1, 2) structural units. At the same time, the Raman peaks broaden, reflecting increased structural disorder induced by Pb doping. In addition, the $A_1$ vibrational mode of edge-sharing $GeTe_{4-n}Ge_n$ ($n$ = 0–3) tetrahedra exhibits a blue shift with increasing Pb content, moving from 134.8 cm$^{-1}$ to 137.5 cm$^{-1}$, indicating changes in the local bonding environment associated with isovalent substitution at Ge sites.

The Raman spectra of the *hcp* phases of Pb-doped GST films reveal the next stage in the crystallization process. Fitted Raman spectra of all films exhibit four characteristic subpeaks. The dominant vibrational modes are attributed to the $A^1_{1g}$, $E^2_g$, and $A^2_{1g}$ modes of $Sb_mTe_3$ ($m$ = 1, 2) units; meanwhile, the $A_1$ vibrational mode of $GeTe_{4-n}Ge_n$ ($n$ = 0–3) units exhibits a blue shift relative to its positions in the *fcc* phase, moving to the 141-145 cm$^{-1}$ range. Moreover, this band, which is intense in undoped GST, becomes significantly suppressed, indicating a reduced contribution of GeTe-related vibrations due to the dominance of $Sb_mTe_3$ ($m$ = 1, 2) units, partly as a result of isovalent substitution of Ge.

Additional EDS elemental mapping of undoped GST and 6.8 at.% Pb-GST films was performed to investigate the elemental distribution across the films (Fig. 2). The results indicate that the elements Ge, Sb, Te, and Pb (in the doped film) are uniformly distributed throughout both the amorphous and *hcp* phases. In the doped film with the highest Pb concentration, no obvious segregation of a secondary phase was detected, suggesting that it is either finely dispersed at the nanoscale or present in only trace amounts.

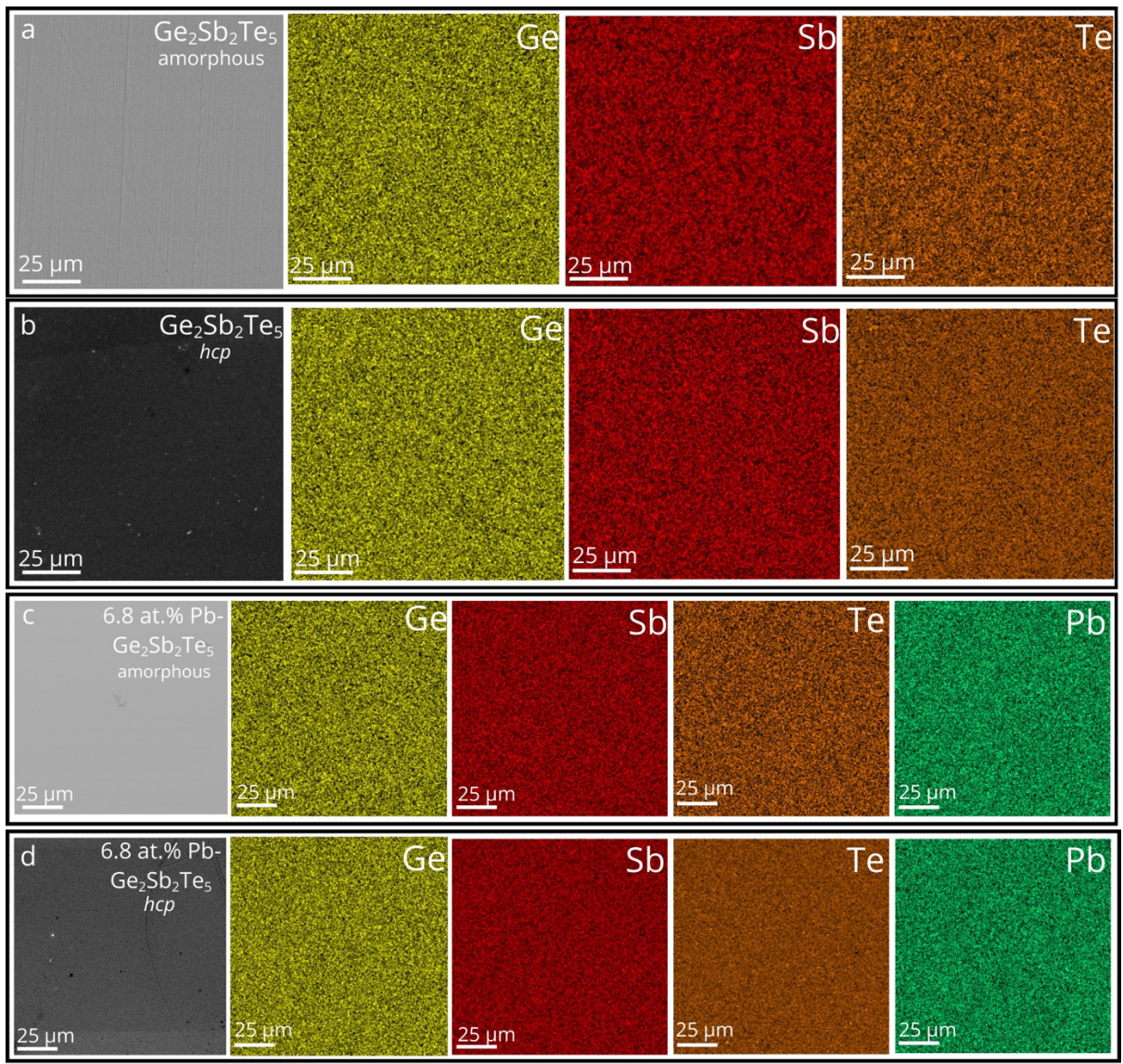

Fig. 2. EDS elemental mapping of GST films: (a) amorphous undoped GST, (b) *hcp* crystallized undoped GST, (c) amorphous 6.8 at. % Pb-doped GST and (d) *hcp* crystallized 6.8 at. %. Pb-doped GST. Elemental distributions are shown as follows: Ge (yellow), Sb (red), Te (orange), and Pb (green).

To further clarify the chemical environment of Pb and its interaction with the GST, XPS measurements were performed on the film with the highest Pb content (6.8 at.%) annealed at 325 °C (Fig. 3). The survey spectrum confirms the presence of Ge, Sb, Te, and Pb in this film. Oxygen and carbon are most likely attributed to residual surface contamination originating from solvents and cleaning agents used during sample preparation, as well as to adventitious carbon and partial surface oxidation of the film during storage, whereas Na and Ca originate from the soda-lime glass substrate. The high-resolution Pb 4*f* spectrum displays two spin–orbit doublets at 137.6 eV (Pb 4$f_{7/2}$) and 142.5 eV (Pb 4$f_{5/2}$), with a spin–orbit splitting of 4.9 eV, corresponding to Pb–Te bonding and Pb in the +2 oxidation state [43, 44]. Additional peaks at 138.7 eV and 143.5 eV indicate the presence of oxidized Pb species [43, 45]. No metallic Pb$^0$ peaks were observed, confirming that Pb exists solely as telluride or oxide in the sample.

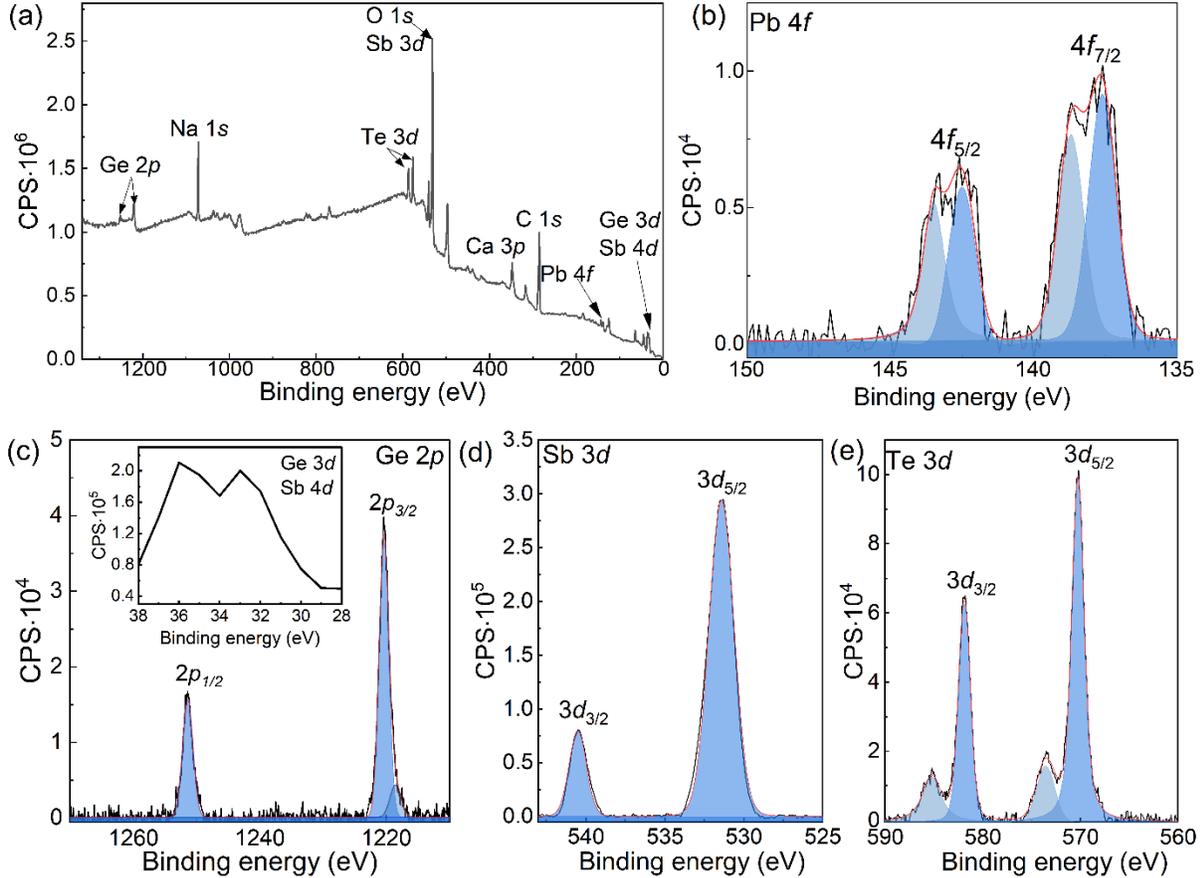

Fig. 3. XPS spectra of 6.8 at.% Pb-doped GST annealed at 325 °C: (a) survey spectrum, (b) high-resolution scan of the Pb 4*f* region, (c) high-resolution scan of the Ge 2*p* region with inset showing the overlapping Ge 3*d* and Sb 4*d* regions, (d) high-resolution scan of the Sb 3*d* region, and (e) high-resolution scan of the Te 3*d* region. High-resolution regions were fitted using Voigt functions.

The Ge 3*d* (29–33 eV [43]) and Sb 4*d* (31–35 eV [46]) regions overlap significantly, complicating reliable deconvolution (see inset Fig.3 (c)). The Ge 2*p* region exhibits a stronger signal, allowing clearer analysis. The spin-orbit doublet corresponding to Ge 2$p_{3/2}$ and Ge 2$p_{1/2}$ is observed at binding energies of 1218.6 eV and 1251.4 eV, respectively, and is attributed to Ge–Te bonding [35, 47]. In addition, the Ge 2$p_{3/2}$ exhibits an additional component at 1220.4 eV, which is assigned to oxidized germanium species [33, 48]. The Sb 3*d* region overlaps with the O 1*s* signal, which complicates peak fitting. Nevertheless, clear peaks are observed at 531.4 eV and 540.5 eV, corresponding to Sb 3$d_{5/2}$ and Sb 3$d_{3/2}$, respectively. These binding energies can be attributed to Sb–Te/Sb–O bonds. [33, 47-49]. The Te 3*d* region exhibits the characteristic doublet, with Te 3$d_{5/2}$ and Te 3$d_{3/2}$ peaks at 571.5 eV and 581.9 eV, respectively, with spin-orbit splitting

10.4 eV [50] consistent with Te$^{2-}$ in tellurides [48]. Minor satellite peaks at slightly higher binding energies correspond to oxidized Te species, indicating minor surface oxidation. The absence of additional peaks confirms that the majority of Te remains in the telluride state. Overall, the XPS analysis of the 6.8 at.% Pb-doped GST film shows the expected chemical states for Ge, Sb, and Te, consistent with typical GST materials. Lead is predominantly in the +2 oxidation state, indicating that it either occupies Ge or Sb sites in the lattice, where it forms Pb–Te bonds, or is present as a minor PbTe phase.

These results suggest that Pb doping significantly influences the structure and bonding of GST films during crystallization. In the amorphous state, increasing Pb content alters chemical bond lengths and angles. In the *fcc* phase, Pb incorporation leads to expanded interplanar spacing due to the larger covalent radius of Pb compared to the host atom. Furthermore, isovalent substitution at Ge sites causes a blue shift of the GeTe-related Raman band. Additionally, 4.8 at.% Pb-GST and 6.8 at.% Pb-GST films show crystallization dominated by Sb$_m$Te$_3$ ($m$ = 1, 2) units, as well as demonstrate a more disordered structure compared to undoped GST and 2.5 at.% Pb–GST. In the *hcp* phase, Pb continues to modify the vibrational dynamics of the GST lattice, enhancing crystallization through the dominance of the Sb$_m$Te$_3$ ($m$ = 1, 2) sublattice. Moreover, Pb incorporation promotes anisotropic lattice changes through structural densification, with minor PbTe formation observed in films with higher Pb content (4.8 and 6.8 at.%). The secondary phase appears to be limited in quantity, with its elements relatively homogeneously distributed across the film surface, showing no obvious local segregation.

## 3.2 Electrical Behavior of Pb-Doped GST Films During Crystallization

GST-based PCMs exhibit a significant contrast between their two stable phases, amorphous and *hcp*. Fig. 4 (a–d) shows the specific electrical resistivity of Pb-doped GST films as a function of temperature. The resistivity changes can be divided into five regions: amorphous state, first phase transition from amorphous to *fcc* ($T_1$), metastable *fcc* phase, second phase transition from *fcc* to *hcp* ($T_2$), and stable *hcp* phase. Undoped GST, used as a reference sample, showed $T_1$ =125 °C, which is consistent with the literature [15]. Increasing the Pb content led to a decrease in the first transition temperature, reaching 93 °C for 4.8 at.% Pb-GST and remaining similar for 6.8 at.% Pb-GST. Regarding the *fcc* to *hcp* transition, undoped GST undergoes this phase transformation at 273 °C. Increasing the Pb content in the films leads to a decrease in the *fcc*-to-*hcp* transition temperature, reaching 240 °C for 6.8 at.% Pb-GST. The phase transition temperatures from amorphous to *fcc* ($T_{1\,d\rho}$) and from *fcc* to *hcp* ($T_{2\,d\rho}$) for all films are also provided in Table S2. In general, crystallization processes require overcoming an activation energy barrier, which is closely related to the average bond strength within the material [51]. In this context, the decrease in transition temperatures observed in Pb-doped GST can be attributed to the substitution of Ge atoms by Pb, which leads to the replacement of stronger Ge–Te bonds (396.7 kJ/mol) with weaker Pb–Te bonds (249.8 kJ/mol) [52]. If Pb also occupies Sb sites, this substitution may additionally replace Sb–Te bonds (277.4 kJ/mol) with Pb–Te bonds, further contributing to the reduction in transition temperatures [52]. The reduction in average bond strength lowers the energy required for atomic rearrangements during crystallization, thereby shifting the phase transitions to lower temperatures. In addition, the more gradual decrease in electrical resistance and the broadened transition regions observed for highly doped samples suggest a modification of crystallization kinetics, possibly associated with changes in nucleation and growth processes [48]. Thus, the observed temperature shifts likely arise from a combination of structural–thermodynamic effects related to bond weakening and kinetic factors influencing phase transformation behavior.

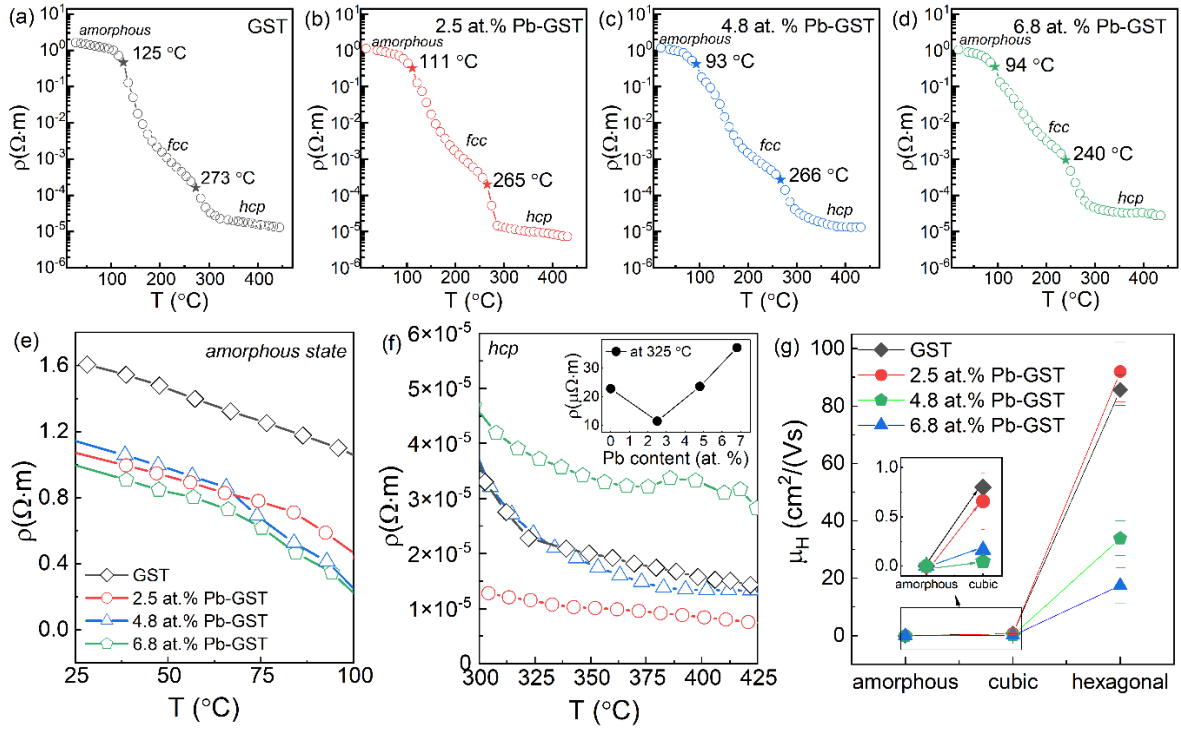

Fig. 4. Temperature-dependent specific electrical resistivity of GST films with phase transition temperatures indicated amorphous → *fcc* and *fcc* → *hcp* for (a) undoped GST, (b) 2.5 at.% Pb-GST, (c) 4.8 at.% Pb-GST, and (d) 6.8 at.% Pb-GST films, measured at a heating rate of 20 °C/min. Specific electrical resistivity of Pb-doped and undoped GST films in (e) their amorphous phase and (f) the *hcp* phase. The inset shows the dependence of resistivity on Pb content at 325 °C. (g) Variation of Hall mobility during crystallization. The inset demonstrates a magnified view of the increase in Hall mobility associated with the transition to the cubic phase. Error bars correspond to the standard deviation of the measured values.

Previous studies have shown that different dopants influence GST crystallization in markedly different ways. For instance, Sn doping was reported to induce only a minor increase in the crystallization temperature (by 5 °C) [53]. In contrast, other studies showed that Sn doping reduced the crystallization temperature by approximately 15 °C while preserving the overall crystal structure and without inducing the formation of a SnTe phase [54]. Bi doping generally preserved the phase-transition sequence of GST but significantly lowered the crystallization temperature (to below 100 °C) and suppressed the second phase transition to below 200 °C with increasing Bi concentration [51]. In contrast, indium doping led to a significant increase in the amorphous-to-crystalline transition temperature, rising from ~160 °C to ~256 °C at high In contents (≈20 at.%) [23]. In this context, Pb incorporation exhibited a distinct behavior, resulting in a reduction of both phase-transition temperatures.

Moreover, it can be observed that Pb doping generally affects the temperature range over which the metastable *fcc* phase is maintained. For undoped GST, the range $\Delta T = T_2 - T_1$ is 148 °C. This range increases to 154 °C for 2.5 at.% Pb–GST and reaches a maximum of 173 °C for 4.8 at.% Pb-GST and then decreases to 146 °C for 6.8 at.% Pb–GST. Notably, 4.8 at.% Pb favors the longest preservation of the metastable phase, indicating that the amorphous-to-*fcc* transition occurs more readily while the *fcc*-to-*hcp* transition temperature is only slightly reduced.

In other words, doping reduces the onset temperature of crystallization while prolonging the stability of the metastable phase. This means that Pb doping provides better control over the metastable phase. Furthermore, the decreased *fcc* to *hcp* transition temperature could potentially lower the power consumption required for the RESET process in phase-change materials.

Fig. 4(e, f) shows the effect of Pb doping on the electrical properties of GST films in both the amorphous and *hcp* phases. In the as-deposited (amorphous) films, the $\rho$ decreases with increasing Pb content. In the *hcp* phase, $\rho$ decreases for 2.5 at.% Pb–GST, but with further increases in Pb concentration, $\rho$ rises. The inset of Fig. 4(f) highlights this trend by showing the specific resistivity values measured at 325 °C, which initially decrease and then increase gradually. This behavior is likely related to changes in the carrier concentration. Previous studies have demonstrated that Pb incorporation into *p*-type semiconductor systems can introduce acceptor levels through substitutional doping. For example, Pb atoms substitute Bi sites in BiCuSeO [55], $Bi_2Te_3$ [56], leading to acceptor behavior, increasing electrical conductivity, and enhancing hole concentration. Similar acceptor-like effects have been reported in Pb-doped $CuSbSe_2$ [57] and $Cu_3SbSe_2$ [58], where Pb partially substitutes Sb atoms, modifying charge transport properties. The transport properties of the hexagonal phase are discussed in more detail in Section 3.3.

Consequently, $\rho$ is governed by the carrier concentration and mobility, as described by

$$\rho = 1/(qn\mu) \quad (1)$$

where $q$ is the carrier charge, $n$-carrier concentration, and $\mu$-carrier drift mobility.

To evaluate the carrier mobility in Pb-GST films, the Hall constant was measured. The Hall measurements revealed clear trends in the transport properties of Pb-doped GST films during crystallization. The Hall mobilities ($\mu_H$) of carriers, extracted *via* the Hall measurements for Pb-doped and undoped GST films at different stages of the crystallization process are shown in Fig. 4(g).

The following represents the results of the experiment in question. The carrier mobility increases significantly as the films crystallize, reaching its maximum for samples in the hexagonal phase. In the amorphous state, the $\mu_H$ is nearly constant. Upon crystallization into the *fcc* phase, $\mu_H$ initially increases, but further Pb doping reduces the mobility. Pb doping in cubic GST likely increases the degree of structural distortion in an already highly disordered lattice. In fact, the crystal structure of cubic GST deviates from the ideal crystalline arrangement. Cubic GST inherently contains about 20% vacancies at the cationic sites [59], and these vacancy-induced structural distortions are responsible for intensive electron scattering and reduced mobility [6]. Embedding Pb in the crystal GST structure likely amplifies these distortions, since despite its electronic similarity to Ge, the atomic size mismatch between Ge and Pb will introduce additional local strain. Consistent with this, a gradual increase in the lattice parameter of the cubic phase is observed with increasing Pb atomic concentration (Table 1). As the structural disorder intensifies, carriers undergo more effective scattering by crystal structure defects and stronger localization due to disrupted orbital overlap [6]. This disorder-induced localization weakens metavalent bonding, reduces electron delocalization, and ultimately lowers the carrier mobility in Pb-doped cubic GST.

In the *hcp* phase, $\mu_H$ rises by about six orders of magnitude compared to the amorphous state, reaching its maximal value for 2.5 at.% Pb–GST film and remaining slightly lower for undoped GST. However, films with higher Pb contents (4.8 and 6.8 at.%) exhibit the lowest mobility values. The elevated mobility of 2.5 at.% Pb–GST film in the *hcp* phase suggests improved crystallinity and reduced disorder. The obtained values are consistent with previously reported data for GST films [6]. The observed decrease in mobility at Pb concentrations above 2.5 at.% can be attributed to excessive impurity incorporation, which introduces additional scattering centers and hinders carrier transport, as will be discussed in Section 3.3

### 3.3 Transport and Thermoelectric Properties of Stable Crystalline Pb-Doped GST Films

The temperature dependences of the thermoelectric properties, including the specific electrical resistivity and Seebeck coefficient, taken within the 300÷700 K range for the *hcp* crystallized GST phase, are presented in Fig. 5. The resistivity linearly increases with growing temperature (Fig. 5(a) that corresponds to the regime of a degenerate semiconductor, where the carrier concentration is *T*-independent. In accordance with expression (1), the temperature evolution of the $\rho(T)$ dependence for the degenerate semiconductors will be primarily governed by the temperature dependence of the carrier mobility. The *T*-linear growth of $\rho$ with increasing temperature can be related to the carriers scattering by acoustic and optical phonons. Above the Debye temperature, the number of phonons increases with *T* and, hence, $\rho$ increases, too.

The sign of the Seebeck coefficient is positive, which is in alignment with the hole conductivity of the samples being studied. Their *p*-type conductivity originates from intrinsic cation (Ge/Sb) vacancies and antisite defects, which create hole carriers [6, 60]. Similarly to the $\rho(T)$, the *S* also linearly increases with growing temperature (Fig. 5(b)).

Temperature dependences of power factor, which were calculated as $PF = S^2/\rho$, for Pb-doped GST films with different doping levels, are shown in Fig. 5 (c). For the undoped GST, the maximum power factor reaches 0.8 mW/(K$^2$·m) at 464 K, while the highest *PF* was obtained for 2.5 at.% Pb-GST reaching 1.3 mW/(K$^2$·m) at 633 K and above. A summary and literature comparison of *PF* values for GST-based films is presented in Table S3. Although several studies have reported on GST-based thermoelectric films, meaningful comparisons require closely matched technological parameters, including the deposition method, film thickness, dopant concentration, and other processing conditions. More broadly, for conventional *p*-type mid-temperature thermoelectric films, typically operating in the 300–600 K range (e.g., Sb$_2$Te$_3$- and GeTe-based systems), and prepared by magnetron sputtering, reported power factors generally range from 0.21 to 4.6 mW/(K$^2$·m) [61-65]. Regarding GST films fabricated by a single method (here, magnetron sputtering), it has been reported that their thermoelectric performance could be enhanced by several factors, including the deposition temperature [63], substrate heating during sputtering [66], and the controlled stacking of individual GeTe and Sb$_2$Te$_3$ layers [67]. In doped GST films, incorporation of elements such as Bi and Sn has been shown to yield *PF* values in the range of approximately 0.55–1.88 mW/(K$^2$·m) [68-70]. Therefore, even when compared with previously reported systems, the Pb-doped Ge$_2$Sb$_2$Te$_5$ films obtained in the present work exhibit competitive *PF* values, exceeding some of the reported films.

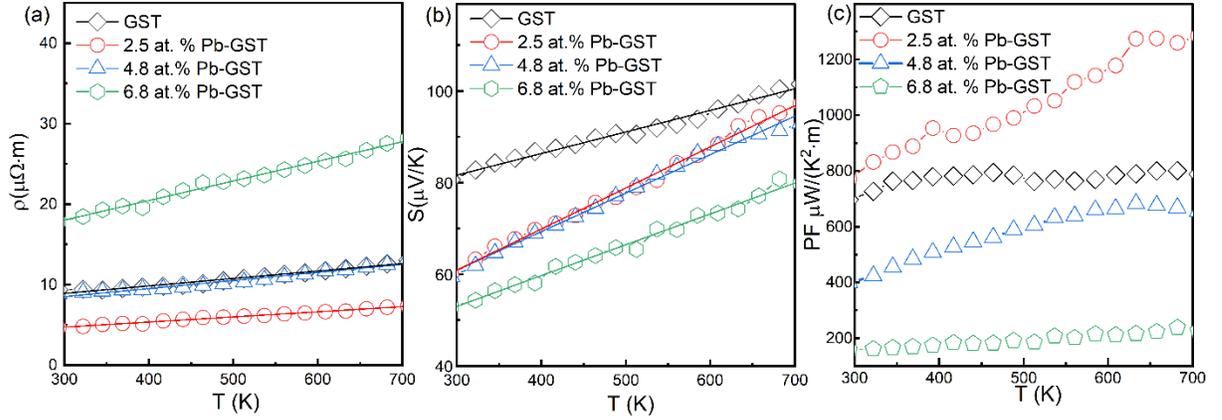

Fig. 5. Thermoelectric properties of *hcp* phase GST and Pb-doped GST films: (a) specific electrical resistivity, (b) Seebeck coefficient, and (c) power factor.

The hole concentration as a function of Pb doping in films annealed at 325 °C, measured at room temperature, is shown in Fig. 6(a). The hole concentration is on the order of $10^{19}$ cm$^{-3}$ and exhibits a linear dependence, increasing with the rising Pb content in the films. According to the Pauling scale, Sb and Pb have the same electronegativity (1.8), while Ge and Pb belong to the same group in the periodic table and commonly exhibit a +2 oxidation state. On this basis, Pb substitution at Ge cation sites can be regarded as isovalent doping and is therefore not expected to significantly change the carrier concentration. In contrast, substitution of Pb at Sb sites leads to acceptor-type doping [57, 58, 71], since Sb is predominantly in the +3 oxidation state in GST. Such substitution effectively generates holes, leading to an increase in carrier concentration. Furthermore, the GST lattice is known to contain a high density of antisite defects, such as Ge$_{Sb}$ и Sb$_{Ge}$ [6, 60], which play a crucial role in determining its electronic properties. Moreover, previous simulation indicated that among the *p*-type defects in ordered GST phases, Ge vacancies (V$_{Ge}$), Ge$_{Sb}$, and Sb$_{Te}$, Ge$_{Sb}$ is the most favorable defect and is therefore expected to play a more important role in *p*-type self-doping than V$_{Ge}$ and Sb$_{Te}$ defects [72]. If Pb substitutes at Ge vacancy sites, the resulting reduction in V$_{Ge}$ would lead to a decrease in hole concentration [73]. However, the experimental results show the opposite trend, indicating the dominance of an alternative mechanism. Since Pb is isovalent with Ge, it is more likely to follow the tendency of *p*-type defects by forming Pb$_{Sb}$ antisite defects. Specifically, Pb incorporation at Sb sites introduces acceptor states, leading to an overall increase in hole concentration, as has also been reported for Pb-doped (Bi$_{0.2}$Sb$_{0.8}$)$_2$Te$_3$, where the Pb acts as a *p*-type dopant generating holes [71]. A similar effect of aliovalent Pb$^{2+}$ doping at Sb$^{3+}$ sites has been reported for CuSbSe$_2$ [57], and Pb introduction has been shown to enhance carrier concentration of *p*-type Cu$_3$SbSe$_4$ [58]. In addition, the formation of Pb$_{Te}$ antisite defects, which act as additional acceptors, may further contribute to the observed increase in the hole concentration, consistent with reports on Sn-doped GST [37]. Thus, based on the analysis of transport properties and defect considerations, the results suggest that, in addition to the substitution of Ge by Pb, the combined effect of Pb substitution at Sb sites and the formation of Pb-related antisite defects may represent favorable pathways responsible for the observed increase in Hall hole concentration in Pb-doped GST films.

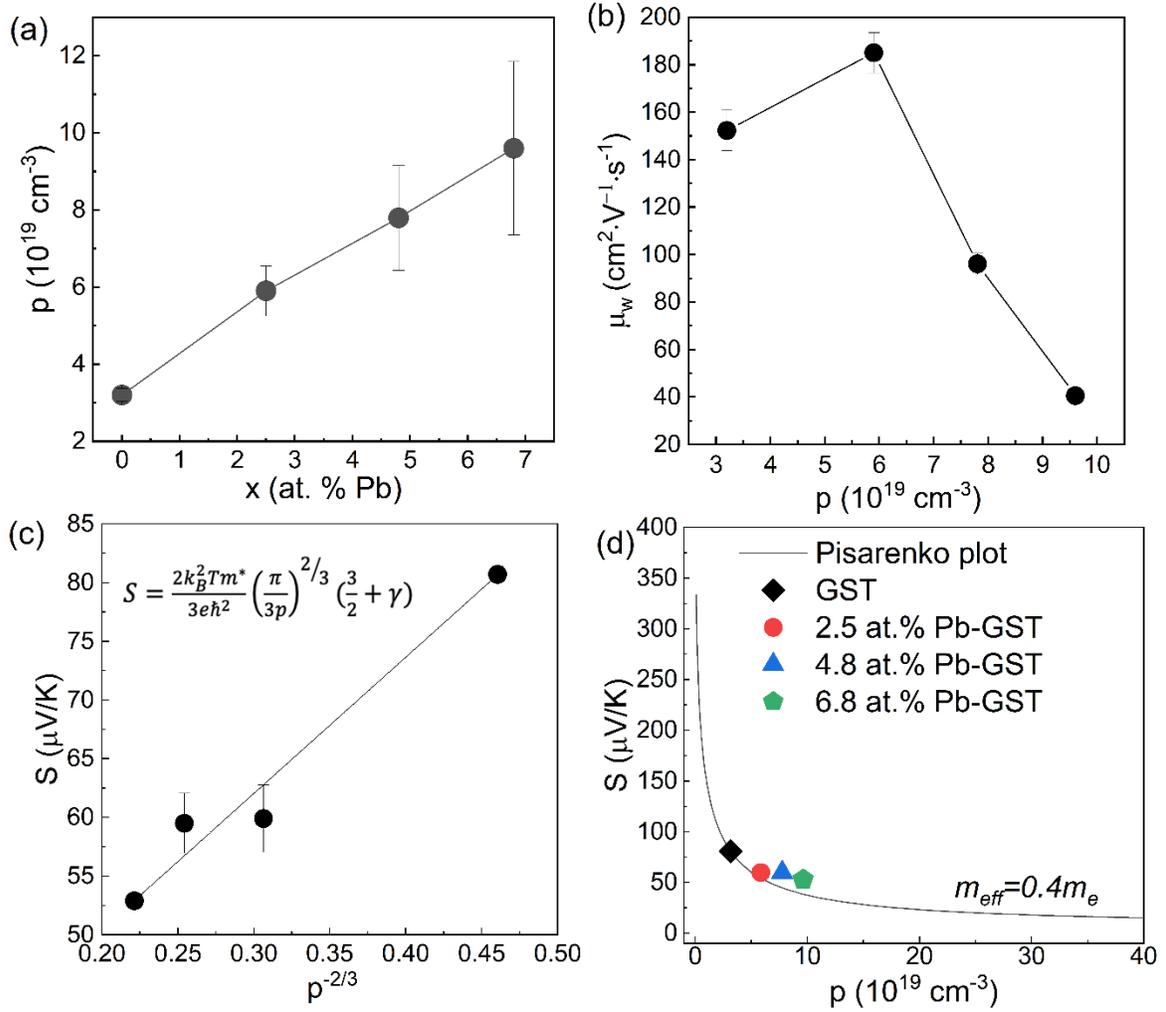

Fig. 6. (a) Carrier concentration of hexagonal phase GST films as a function of Pb atomic content. Error bars correspond to the standard deviation of the measured values. (b) Dependence of the weighted mobility $\mu_w$ of hexagonal phase crystallized films on the carrier concentration at room temperature. Error bars represent the propagated uncertainty from the Seebeck coefficient and resistivity measurements. (c) Dependence of the Seebeck coefficient on $p^{-2/3}$. (d) Pisarenko plot showing the relationship between hole concentration and Seebeck coefficient at room temperature. The points correspond to the experimental values for GST films with different Pb atomic concentrations.

The weighted mobility ($\mu_w$) of the *hcp* phase, which is related to the drift mobility ($\mu$), was also analyzed. We employed the $\mu_w$, which considers the electron mobility scaled by the density of electronic states [74]. Within the free-electron model, assuming a constant mean free path, the $\mu_w$ can be expressed as a simple analytic function of the electrical conductivity and Seebeck coefficient, closely approximating the Drude–Sommerfeld model (within ~3% for |S| > 20 μV /K) [74]. $\mu_w$ provides an estimate of the intrinsic drift mobility of charge carriers and can be expressed as a function of the Seebeck coefficient and the electrical resistivity, as defined below:

$$\mu_w = \frac{3h^3}{8\pi e (2m_e k_B T)^{3/2} \rho} \left[ \frac{exp\left[\frac{|S|}{k_B/e} - 2\right]}{1 + exp\left[-5\left(\frac{|S|}{k_B/e} - 1\right)\right]} + \left[ \frac{\frac{3}{\pi^2} \frac{|S|}{k_B/e}}{1 + exp\left[5\left(\frac{|S|}{k_B/e} - 1\right)\right]} \right] \right] \quad (2)$$

where $m_e$ is the free electron mass.

Generally, a relationship between the $\mu_w$ and $\mu$ can be expressed by:

$$\mu_w = \mu \left(\frac{m^*}{m_e}\right)^{3/2} \quad (3)$$

The results of $\mu_w$ are presented in Fig. 6(b) and show a clear maximum at 2.5 at.% Pb, in agreement with the trend observed for the Hall mobility (Fig. 4(g)). According to solid-state physics, lattice distortion, defect formation, and increased carrier concentration are generally expected to reduce carrier mobility due to enhanced scattering [75]. This behavior is indeed observed in GST films at Pb concentrations of 4.8 at.% and higher, where mobility decreases as defect and impurity scattering become dominant. However, at low Pb concentrations, the observed mobility enhancement can be rationalized by microstructural and defect-chemistry effects. Previous studies have reported that doping can induce changes in film texture in GST-based systems (e.g., Sn-doped $Ge_2Sb_2Te_5$), leading to anisotropic grain growth along specific crystallographic directions [54]. Such texturing can modify the grain microstructure and reduce carrier scattering at grain boundaries, thereby enhancing mobility. With further Pb addition $\mu_w$ decreases, while the hole concentration continues to rise. It appears that Pb concentrations above 2.5 at.% introduce a larger number of charged scattering centers and additional local strain, which enhance momentum relaxation and reduce the drift mobility. Furthermore, Pb incorporation can modify the band curvature or create localized electronic states; in particular, at higher concentrations, where the formation of a PbTe secondary phase is observed. In terms of solid-state physics, the minor PbTe phase is expected to contribute to charge-carrier scattering, which manifests as a reduction in carrier mobility, particularly for Pb-GST samples with 4.8 and 6.8 at.% Pb. The thermoelectric behavior, as reflected in the monotonic decrease of the Seebeck coefficient with increasing doping level, is also consistent with the corresponding increase in carrier concentration. PbTe is a narrow-bandgap semiconductor, similar to GST, with a comparable bandgap width [6, 52]. Accordingly, in samples with Pb concentrations of 4.8 at.% and 6.8 at.%, where a secondary PbTe phase is present, one may expect effects associated with electron energy filtering as well as enhanced scattering of charge carriers and phonons. Among these mechanisms, charge carrier scattering appears to be the dominant effect, as evidenced by the observed reduction in carrier mobility. Although Pb doping reduces carrier mobility, it remains challenging to disentangle the individual contributions of doping-induced disorder and secondary-phase formation to the overall scattering. Hence, PbTe inclusions are expected to act as phonon-scattering centers; however, a reliable quantitative evaluation of this effect is difficult. Overall, 2.5 at.% Pb represents a doping optimum: Pb addition increases the carrier concentration and can improve transport quality, whereas higher doping enhances scattering and structural disorder.

Seebeck coefficient is determined by both carrier concentration and carrier effective mass:

$$S = \frac{2k_B^2 T m^*}{3e\hbar^2} \left(\frac{\pi}{3p}\right)^{2/3} \left(\frac{3}{2} + \gamma\right) \quad (4)$$

where $\hbar$ the reduced Planck constant, $p$-hole concentration, $m^*$ is the density-of-state effective mass of electrons, and $\gamma$ is the scattering factor.

In accordance with expression (4), the $S-p^{-2/3}$ dependence is expected to be linear. As shown in Fig. 6 (c), the data indeed remain approximately linear.

The Pisarenko plot is a well-known tool used in thermoelectric research to analyze the relationship between the Seebeck coefficient and the carrier concentration in semiconducting materials. Moreover, it has been shown that this remains true even beyond the non-degenerate regime [76].

In the case of Pb-doped GST films (Fig. 6(d)), the overall agreement with the Pisarenko curve suggests that the Pb incorporation does not drastically change the band structure at low doping levels, with the primary effects arising from changes in carrier concentration and scattering. Although secondary phase formation is present at higher Pb contents and is uniformly distributed across the films, small deviations observed for Pb concentrations of 4.8 and 6.8 at.% are more cautiously attributed to additional scattering mechanisms and local energy filtering associated with the secondary phase, rather than to pronounced band-structure changes.

## 4. Conclusion

The effect of Pb doping on the crystallization process and thermoelectric behavior of GST films was systematically investigated. Low Pb content (2.5 at.%) introduces only slight local distortions without significantly modifying the crystalline lattice, whereas higher concentrations (4.8 and 6.8 at.%) lead to pronounced disorder and an enhanced crystallization pathway dominated by $Sb_mTe_3$ ($m$ = 1, 2) units. Pb doping improves control over the metastable cubic phase, lowers both the crystallization temperature for amorphous-to-*fcc* transformation and the temperature for structural *fcc*-to-*hcp* transition, indicating a potential reduction in the energy required for phase switching. The incorporation of Pb atoms increases the hole concentration, likely due to Pb substitution not only at Ge sites but also at Sb sites, as well as the formation of Pb-related antisite defects. At higher Pb contents, increased structural disorder and scattering on impurities reduce carrier mobility while increasing hole concentrations. Electrical measurements indicate that crystalline 2.5 at.% Pb-GST film provides the most favorable thermoelectric performance, with optimized resistivity and mobility in the *hcp* phase. The Pisarenko plot analysis suggests that 2.5 at.% Pb primarily modulates the carrier concentration without significantly affecting the band structure. Overall, a Pb doping concentration of around 2.5 at.% and below 4.8 at. %, represents an optimal doping window, balancing structural stability with improved charge-transport. In contrast, excessive Pb incorporation leads to carrier mobility degradation, which is more cautiously attributed to additional scattering mechanisms and local energy filtering associated with the formation of a secondary phase. These results demonstrate that Pb is an effective dopant for tuning the crystallization behavior and electronic properties of GST, offering promising opportunities for next-generation energy technologies and on-chip thermoelectric generators.

# Supplementary Materials for

# Effect of Pb doping on the crystallization process and thermoelectric properties of Ge$_2$Sb$_2$Te$_5$ phase change material

M. Zhezhu [1*], A. Vasil'ev [1], M. Yaprintsev [2], A. Musayelyan [3], E. Pilyuk [2], O. Ivanov [2]

[1] A.B. Nalbandyan Institute of Chemical Physics NAS RA, Yerevan 0014, Armenia
[2] Belgorod State University, Belgorod 308000, Russia
[3] Institute of Radiophysics and Electronics NAS RA, Armenia

*Author to whom correspondence should be addressed: marina.zhezhu@ichph.sci.am

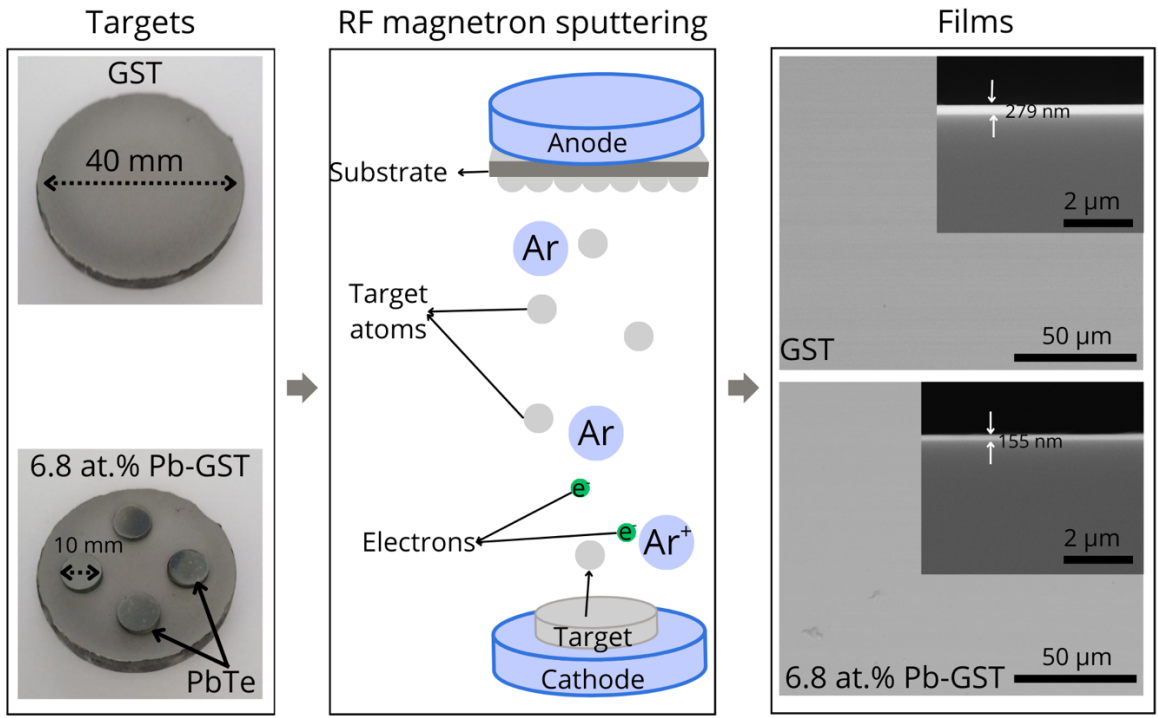

Fig. S1. Schematic representation of film sputtering, showing the targets used, the deposition process, and SEM analysis of the obtained films (pure GST and GST with the maximum Pb content).

Table S1. Gaussian Fitted Raman peak positions of the films

| Measurement temperature | GST | 2.5 at. % Pb-GST | 4.8 at. % Pb-GST | 6.8 at. % Pb-GST |
|---|---|---|---|---|
| | Peak position, cm$^{-1}$ | | | |
| 25 °C | 107.5 | 115.9 | - | - |
| | 144.6 | 135.5 | 138.1 | 140.8 |
| | 158.8 | 145.1 | 155.8 | 155.1 |

| 225 °C | - | - | 63.1 | 63.5 |
| --- | --- | --- | --- | --- |
| | 115.0 | 118.5 | 105.1 | 109.6 |
| | 134.8 | 137.0 | 140.2 | 137.5 |
| | 151.5 | 150.9 | 159.3 | 157.0 |
| 325 °C | 61.5 | 65.0 | 65.1 | 64.6 |
| | 105.1 | 109.1 | 107.4 | 107.3 |
| | 141.1 | 141.6 | 141.0 | 145.4 |
| | 157.8 | 162.2 | 160.3 | 160.6 |

Table S2. Temperatures of phase transitions: from amorphous to *fcc* ($T_{1\,d\rho}$) and from *fcc* to *hcp* ($T_{2\,d\rho}$).

| | GST | 2.5 at.% Pb-GST | 4.8 at.% Pb-GST | 6.8 at.% Pb-GST |
| --- | --- | --- | --- | --- |
| $T_{1\,d\rho}$, °C | 125 | 111 | 93 | 94 |
| $T_{2\,d\rho}$, °C | 273 | 265 | 266 | 240 |

Table S3. Summary of *PF* values for mid-temperature thermoelectric films (GeTe- and $Sb_2Te_3$-based systems), as well as optimized and doped crystalline GST films prepared by magnetron sputtering.

| Film | Maximum *PF* | Reference |
| --- | --- | --- |
| $Sb_2Te_3$ | 0.27 mW/(K$^2$·m) at 483 K | S1 |
| Cu-doped $Sb_2Te_3$ | 0.38 mW/(K$^2$·m) at 483 K | |
| Ag-doped $Sb_2Te_3$ | 4.6 mW/(K$^2$·m) at 373 K | S2 |
| $Bi_{0.5}Sb_{1.5}Te_3$ | 0.21 mW/(K$^2$·m) at RT | S3 |
| GeTe modulated by working pressure | 2.0 mW/(K$^2$·m) at 563 K | S4 |
| GeTe | 0.85 mW/(K$^2$·m) at 300 K; 1.89 mW/(K$^2$·m) at 575 K | S5 |
| In-doped GeTe | 1.41 mW/(K$^2$·m) at 300 K; 2.92 mW/(K$^2$·m) at 575 K | |
| GST films optimized by controlling the deposition temperature | 1.1 mW/(K$^2$·m) at RT | S3 |
| GST films optimized by substrate heating during sputtering | 0.77 mW/(K$^2$·m) at RT | S6 |
| GeTe/$Sb_2Te_3$ multilayer films with controlled stacking periods of the individual GeTe and $Sb_2Te_3$ layers | 1.088 mW/(K$^2$·m) at 473 K | S7 |
| Sn-doped $Ge_2Sb_2Te_5$ | 1.88 mW/(K$^2$·m) at 723 K | S8 |
| Sn-doped $Ge_2Sb_2Te_5$ | 1.7 mW/(K$^2$·m) at 450 K | S9 |
| Bi-doped $Ge_2Sb_2Te_5$ | 0.55 mW/(K$^2$·m) at 673 K | S10 |
| Pb-doped $Ge_2Sb_2Te_5$ | 1.3 mW/(K$^2$·m) at 633 K | This work |